\documentclass{article}\usepackage{amsfonts}\begin{document}
\title{ The impact of a random metric upon a diffusing particle }

\author{Z. Haba \\
Institute of Theoretical Physics, University of Wroclaw,\\ 50-204
Wroclaw, Plac Maxa Borna 9, Poland}\maketitle

\begin{abstract} We show that if the singularity of the covariance of the random
metric is  $\vert {\bf x}\vert ^{-2\gamma}$ then the mean value of
the fourth power of the distance achieved in time $t$ by a
diffusing particle behaves as $t^{2(1-\gamma)}$ for a small
$t$.\end{abstract}
\section{Introduction}
An interest in a random diffusion appears in a theory of complex
systems as well as in quantum gravity. In the first case one
considers fluctuating diffusivity ( see
\cite{new}\cite{metzler3}\cite{metzler4}) or random walks with
modified jump probabilities (an overview in ref. \cite{klafter}).
The random metric in quantum field theories is studied  in an
attempt to construct a quantum field theory of all interactions.
There is an old suggestion associated with the names of Landau and
Pauli that the quantization of gravity can change the short
distance behaviour of quantum field propagators allowing to ease
the quantum field theory ultraviolet  problems (for some recent
investigations,see \cite{wilczek}\cite{petrykowski}) . The
propagator is defined as the kernel of the inverse  of the second
order differential operator. In the proper time representation the
propagator is expressed by the heat kernel (as an integral over
time). Then, a change of the short time behaviour of the heat
kernel in a quantized metric leads to a modification of the short
distance behaviour of the propagators. The problem has been
studied numerically in causal triangulation approximation to
quantum gravity \cite{jurk}\cite{jurk2}\cite{jurk3}, in Horava's
gravity at Lifshitz point \cite{hor}\cite{hor2} and in Liouville
gravity \cite{liou}. These studies support the suggestion that
quantum field theory in a quantized metric behaves like a field
theory in lower dimensions (for recent reviews see
\cite{carlip}\cite{carlip2} and references cited there). In a
diffusion theory a random metric in the generator of diffusion can
be interpreted as a random diffusion matrix. One can consider the
additional randomness as a result of a diffusion on irregular
(e.g., fractal) structures
\cite{carlip}\cite{dunne}\cite{proc}\cite{bar}. It is known that
during such a diffusion the short time behaviour of the transition
function is changing. In particular, the mean distance achieved by
a diffusing particle at small time is increasing faster  than in
the regular case. On the basis of this time behaviour one can
define the random walker dimension which is related to the
Hausdorf dimension and to the spectral dimension of the given
fractal structure \cite{dunne}\cite{carlip}. In relation to
quantum gravity it has been suggested already by J.A. Wheeler (see
a discussion in \cite{carlip2}\cite{smolin}) that a "foamy"
structure of space-time at short distances can lead to an
anomalous (more regular) behaviour of matter field propagators at
short distances. As the field propagator is an integral over the
time of the heat kernel the modified short time behaviour of the
heat kernel implies a change of the short distance behaviour of
the propagator. In particular, the results on a diffusion on
fractals \cite{bar} imply that the field propagators on fractal
spaces are more regular.

In this paper we discuss a space-dependent  random diffusion
matrix mimicking the random metric of quantum (Euclidean) gravity.
As long as the random metric has a regular covariance there is no
effect upon the mean distance of a diffusing particle achieved for
a small time $t$.
  We show that
 the mean fourth power of the distance achieved  by a diffusing particle behaves
as $t^{2(1-\gamma)}$ if the covariance of the metric has the
singularity $\vert{\bf x}\vert^{-2\gamma}$( this short time
behaviour would be $t^{2}$ for a diffusion on regular structures;
we can treat rigorously the singularity with
$0\leq\gamma<\frac{1}{2}$). A transition function for a random
 singular metric requires a renormalization. After the renormalization
 the small time behaviour is changed as can already be seen by a
 scaling argument. We define the metric as a square of a tetrad
(vierbein) in order to achieve positivity of the regular metric.
Then, we consider the  tetrad as a Gaussian process with a
singular covariance. We define the singular metric as a Wick
square of the tetrad . In such a case  a renormalization of the
transition function is needed when we go from a regular to a
singular metric. This renormalization is the origin of the
modified short time behaviour. The final effect is the same as the
diffusion on the fractal structures
\cite{bar}\cite{dunne}\cite{carlip} indirectly confirming that a
manifold equipped with a singular random metric  may play the same
role as a fractal space-time.

 In \cite{habaplb} and
\cite{habajpa} we have discussed  a modified small time and short
distance behaviour by means of a scaling transformation in the
functional integral. The argument using the scaling transformation
does not take into account that a definition of correlations of
quantum fields requires normal ordering of fields and coupling
constants renormalization. The eventual scaling should be used in
conjunction  with the renormalization group methods to establish
the scale invariance at the renormalization group fixed point. In
this paper we show by means of explicit calculations that the
singular metric can indeed change the small time behaviour of the
diffusing particle.
\section{Diffusion in a random metric}
In Euclidean quantum field theory correlation functions of the
Euclidean  field $\phi$ defined on a multidimensional Riemannian
manifold (of $D=n+d$ dimensions) are determined by a formal
Gaussian measure
\begin{displaymath}
d\phi\exp (-\frac{\sigma^{2}}{2}\int d^{D}\xi {\cal L}),
\end{displaymath}
with
\begin{displaymath}
{\cal L}=\sqrt{g}g_{AB}\partial^{A}\phi\partial^{B}\phi,
\end{displaymath}
where $g_{AB} $ is the Riemannian metric,$g=\det{g_{AB}}$ and
$\partial^{A}=\frac{\partial}{\partial \xi_{A}}$. The two-point
correlation function (the propagator)
 is the kernel of an inverse of the Laplace-Beltrami operator
$\tilde{{\cal A}}$
\begin{displaymath} \tilde{\cal
A}=\frac{\sigma^{2}}{2}\Big(g^{AB}\frac{\partial^{2}}{\partial
\xi^{A}\partial\xi^{B}}+\Gamma^{A}\frac{\partial}{\partial
\xi^{A}}\Big)
\end{displaymath}
where $\Gamma^{A}$ is the Christoffel symbol (expressed by
derivatives of the metric) and we introduced a diffusion constant
$\sigma^{2}$. In a generator of a diffusion the metric plays the
role of the space-dependent diffusion matrix and $\Gamma$ is a
drift. In quantum gravity we still need to average the scalar
field correlations with respect to a measure over the metric. In a
perturbative expansion the first approximation to this measure is
a Gaussian one with the covariance $ (-\triangle)^{-1}$ where
$\triangle $ is the Laplacian on $\mathbb{R}^{n+d}$.

 We consider a simplified version of the Laplace-Beltrami operator
$\tilde{{\cal A}}$ on $\mathbb{R}^{n+d}$ with coordinates
$\xi=({\bf x},{\bf X})$ where ${\bf x}\in \mathbb{R}^{n}$ and
${\bf X}\in \mathbb{R}^{d}$. We assume that the metric
$g^{AB}=(\delta^{ij},g^{\mu\nu}({\bf x}))$ has a block form. It is
flat on a submanifold described by the coordinates ${\bf x}$ and
depends only on ${\bf x}$ when restricted to the (physical)
submanifold described by the coordinates ${\bf X}$. A partial
justification of such a choice could come from Horava gravity
\cite{hor}\cite{hor2} (motivated by condensed state theory) where
the ${\bf x}$ and ${\bf X}$ coordinates can scale in a different
way so that finally the ${\bf X}$ dependence of the metric is
negligible. Under this assumption the Laplace-Beltrami operator
takes the form
\begin{equation}
\tilde{{\cal A}}=\frac{\sigma^{2}}{2}\Big(\nabla_{{\bf x}}^{2}+
(\nabla_{{\bf x}}\ln \det g) \nabla_{{\bf x}}+ g^{\mu\nu}({\bf
x})\partial_{\mu}\partial_{\nu}\Big),
\end{equation}where $\det g=\det(g_{\mu\nu}) $ is the determinant
of the matrix $g_{AB}$ and
\begin{displaymath}
\partial_{\mu}=\frac{\partial}{\partial X^{\mu}}.
\end{displaymath}
We further simplify the operator $\tilde{{\cal A}}$ neglecting the
drift term $\nabla_{{\bf x}}\ln \det g$. We arrive to the
diffusion generator
\begin{equation}
{\cal A}=\frac{\sigma^{2}}{2}\Big(\nabla_{{\bf
x}}^{2}+g^{\mu\nu}({\bf x})\partial_{\mu}\partial_{\nu}\Big),
\end{equation}
We study  the short time behaviour of the diffusion generated by
${\cal A}$ (2) with a singular random diffusion metric $g$. The
effect of a drift has been studied in \cite{bouchaud}\cite{komor}
( and in references cited there) for a regular random drift. It
has been shown that a random drift can lead to a superdiffusion at
large time but it does not change the short time behaviour. We
shall give some arguments further on that this is the singular
diffusion matrix $g$ (rather than a drift) that leads to an
anomalous diffusion at small time.

In order to achieve the positive definiteness of the metric let us
represent it by the tetrads (vierbeins) $e^{\mu}_{a}$
\begin{displaymath}
g^{\mu\nu}({\bf x})=e^{\mu}_{a}({\bf x})e^{\nu}_{a}({\bf x}),
\end{displaymath}where initially we assume  that $e^{\mu}_{a}({\bf x})$
are regular functions of ${\bf x}$.   We define the stochastic
process (as in \cite{habajpa})
\begin{equation}
{\bf x}_{t}({\bf x})={\bf x}+\sigma{\bf b}_{t},
\end{equation}
\begin{equation}
X^{\mu}_{t}({\bf X})=X^{\mu}+\sigma\int_{0}^{t}e^{\mu}_{a}({\bf
x}_{s})dB_{s}^{a},
\end{equation}
where $({\bf b}_{t},{\bf B}_{t})$ is the Brownian motion on
$\mathbb{R}^{n+d}$ ,i.e., the Gaussian process with mean zero and
the covariance
\begin{displaymath}
\mathbb{E}[b^{j}_{t}b_{s}^{l}]=min(t,s)\delta^{jl}
\end{displaymath}
 (and similarly for ${\bf B}_{t}$).

If there is a drift (as in eq.(1)) then ${\bf x}_{t}({\bf x})$
satisfies an integral equation \begin{displaymath} {\bf
x}_{t}({\bf x})={\bf x}+\int_{0}^{t}\nabla_{{\bf x}}\ln \det
g({\bf x}_{s})ds +\sigma{\bf b}_{t}.
\end{displaymath}
There is an additional time integral on the rhs od this equation
suggesting that for a small time the drift can be neglected.

 It is well-known \cite{gikhman} that the transition function
$P_{t}({\bf x},{\bf X};{\bf y},{\bf Y})$ of the diffusion process
defines a semi-group $\exp(t{\cal A})$. So acting on a function
$\psi$

   \begin{equation}
\Big(\exp(t{\cal A})\psi\Big)\Big({\bf x},{\bf X}\Big)=\int d{\bf
y}d{\bf Y}P_{t}({\bf x},{\bf X};{\bf y},{\bf Y})\psi({\bf y},{\bf
Y})=\mathbb{E}\Big[\psi({\bf x}_{t},{\bf X}_{t})\Big].
\end{equation}
It follows from equation (5) that $P_{t}({\bf x},{\bf X};{\bf
y},{\bf Y})$ is the kernel $K$ of the operator $\exp(t{\cal A})$.
This kernel can be expressed as \begin{equation} K_{t}({\bf
x},{\bf X};{\bf y},{\bf Y})=\mathbb{E}\Big[\delta({\bf y}-{\bf
x}_{t}({\bf x}))\delta({\bf Y}-{\bf X}_{t}({\bf X}))\Big]
\end{equation}
or after the Fourier representation of $\delta$-functions
\begin{equation}\begin{array}{l}
K_{t}({\bf x},{\bf X};{\bf y},{\bf Y})=(2\pi)^{-n-d}\int d{\bf
p}d{\bf P}\cr \mathbb{E}\Big[\exp(i{\bf p}({\bf y}-{\bf
x}_{t}({\bf x})))\exp(i{\bf P}({\bf Y}-{\bf X}_{t}({\bf
X})))\Big].
\end{array}\end{equation}
We wish to calculate an average  $<K_{t}>$ of the diffusion kernel
$K_{t}$ over the metric. For a Gaussian random field $F$
\begin{displaymath}
<\exp F>=\exp(<F>+\frac{1}{2}<F^{2}>)
\end{displaymath}
For non-Gaussian fields these are the first terms of the cumulant
expansion.

We assume that $e^{\mu}_{a}=\delta^{\mu}_{a}+\epsilon^{\mu}_{a}$
is the Gaussian random field with mean $\delta^{\mu}_{a}$ and the
covariance
\begin{equation}
\Big<\epsilon^{\mu}_{a}({\bf x})\epsilon^{\nu}_{c}({\bf
y})\Big>=\kappa\delta^{\mu a;\nu c}G({\bf x}-{\bf y}).
\end{equation} (for simplicity of calculations we choose the matrix on the rhs of eq.(8) in the form
$ \delta^{\mu a;\nu c}= \delta^{\mu \nu}\delta_{ac}$). We obtain a
random perturbation of the Euclidean metric. In perturbative
quantum gravity $G$ is the two-point correlation function of the
graviton.$G({\bf x}-{\bf y})\simeq \vert {\bf x}-{\bf
y})\vert^{-2\gamma}$ ,where $2\gamma=n-2$ if the graviton moves in
the $\mathbb{R}^{n}$ space and $2\gamma=n+d-2$ if the graviton
lives in the $\mathbb{R}^{n+d}$ space. In a rigorous formulation
we must restrict ourselves to $2\gamma<1$.There are some results
based on computer simulations \cite{jurk}
\cite{jurk2}\cite{jurk3}(for $n+d=4$) suggesting that because of
the graviton self-interaction $\gamma<1$. For the semigroup
$\exp(t{\cal A})$ generated by the operator ${\cal A}$ (2) ${\bf
x}_{t}$ (3) does not depend on the metric whereas ${\bf X}_{t}$
being linear in $\epsilon$ is Gaussian. Then, the mean value of
$K_{t}$ in the Gaussian field $e$ is
\begin{equation}\begin{array}{l} \Big<K_{t}({\bf x},{\bf X};{\bf
y},{\bf Y})\Big>=(2\pi)^{-d}\int d{\bf P}\exp(i{\bf P}({\bf
Y}-{\bf X}))\cr\mathbb{E}\Big[\delta({\bf y}-{\bf x}_{t}({\bf
x}))\exp\Big(-i\sigma {\bf P}^{a}{\bf
B}_{t}^{a}-\frac{\sigma^{2}}{2} \Big<(\int_{0}^{t}
P_{\mu}\epsilon^{\mu}_{a}({\bf
x}_{s})dB^{a}_{s})^{2}\Big>\Big)\Big].
\end{array}\end{equation}
For the general Riemannian model (1) the process ${\bf x}_{t}$
with a non-linear drift $\nabla\ln \det g$ is non-Gaussian. A
calculation of expectation values of $\exp(t\tilde{{\cal A}})$
could be performed only in an approximate way, e.g., in cumulant
expansion. As discussed at eq.(1) we rely on the assumption that a
process with a constant diffusion and a random drift has the same
short time behaviour as the Brownian motion ${\bf b}_{t}$. In such
a case the estimates obtained for $<\exp(t{\cal A})>$ will be
valid also for $<\exp(t\tilde{{\cal A}})>$.

The last term in eq.(9) is
\begin{equation}
\exp(-\frac{\sigma^{2}}{2}<(P_{\mu}Q_{t}^{\mu})^{2}>)\end{equation}
where
\begin{equation}
P_{\mu}Q_{t}^{\mu}\equiv P_{\mu}\int_{0}^{t}
\epsilon^{\mu}_{a}({\bf x}_{s})dB^{a}_{s}.
\end{equation}
From the Ito formula \cite{gikhman}\cite{simon}
\begin{equation}\begin{array}{l}
d({\bf
PQ}_{t})^{2}=2P_{\mu}Q^{\mu}_{t}P_{\nu}dQ^{\nu}_{t}+P_{\mu}dQ^{\mu}_{t}P_{\nu}dQ^{\nu}_{t}\cr
=2P_{\mu}\int_{0}^{t}\epsilon^{\mu}_{a}({\bf
x}_{s})dB_{s}^{a}P_{\nu}\epsilon^{\nu}_{c}({\bf
x}_{t})dB_{t}^{c}+P_{\mu}P_{\nu}\epsilon^{\nu}_{a}({\bf
x}_{t})\epsilon^{\mu}_{a}({\bf x}_{t})dt
.\end{array}\end{equation} Hence, integrating eq.(12)
\begin{equation}\begin{array}{l}
({\bf PQ}_{t})^{2}=2P_{\mu}\int_{0}^{t}\epsilon^{\mu}_{a}({\bf
x}_{s})dB_{s}^{a}\int_{0}^{s}P_{\nu}\epsilon^{\nu}_{c}({\bf
x}_{s^{\prime}})dB^{c}_{s^{\prime}}\cr+P_{\mu}P_{\nu}\int_{0}^{t}\epsilon^{\mu}_{a}({\bf
x}_{s})\epsilon^{\nu}_{a}({\bf x}_{s})ds
.\end{array}\end{equation} The formula (13) is sometimes
considered as a definition of the double stochastic integral
\cite{berger}; it appears in quantum electrodynamics
\cite{albeverio}\cite{betz}.

 At the beginning we treat $\epsilon^{\mu}_{a}$ as a regularized random field. In such a case the
correlation function $G$ (8) is a regular function. Now, we remove
the regularization admitting singular $G$. In order to make ${\cal
A}\psi$  a well-defined random field we need the normal ordering
of ${\cal A}$
\begin{equation}
:{\cal A}:=\frac{1}{2}\Big(\nabla_{{\bf x}}^{2}+:g^{\mu\nu}:({\bf
x})\partial_{\mu}\partial_{\nu}\Big),
\end{equation}
where
\begin{displaymath}
:g^{\mu\nu}:=:e^{\mu}_{a}({\bf x})e^{\nu}_{a}({\bf
x}):=e^{\mu}_{a}({\bf x})e^{\nu}_{a}({\bf
x})-<\epsilon^{\mu}_{a}({\bf x})\epsilon^{\nu}_{a}({\bf x})>.
\end{displaymath}
After the normal ordering in eq.(10)
\begin{equation}
<({\bf PQ}_{t})^{2}>\rightarrow <({\bf PQ}_{t})^{2}>-{\bf
P}^{2}t\kappa G(0)d\equiv L_{t}.
\end{equation}
We treat the renormalized kernel $K^{R}_{t}$ in eqs.(9)-(10)as a
generalized function acting on  regular functions $\psi$ ( then
$:{\cal A}:\psi$ is a well-defined random field). After averaging
over the translation invariant random field $e^{\mu}_{a}({\bf x})$
and the renormalization (15) we can write $<\exp (t:{\cal
A}:)\psi>$ in terms of Fourier transforms as

\begin{equation}\begin{array}{l} \int d{\bf y}d{\bf Y}\Big<K^{R}_{t}({\bf x},{\bf X};{\bf
    y},{\bf Y})\Big>\psi({\bf y},{\bf Y})\cr=(2\pi)^{-d}\int d{\bf
    P}d{\bf K}d{\bf k}d{\bf Y}\exp(i{\bf
    P}({\bf Y}-{\bf X} ))\tilde{\psi}({\bf k},{\bf K})\cr
\mathbb{E}\Big[\exp\Big(-\frac{\sigma^{2}}{2} {\bf
    P}^{2}L_{t}\Big)\exp(-i{\bf KY}-i\sigma {\bf P}^{a}{\bf
B}_{t}^{a}-i{\bf k}{\bf x}_{t}({\bf x}))\Big],
\end{array}\end{equation}
 where \begin{equation}\begin{array}{l}
L_{t}=2\int_{0}^{t}dB_{s}^{a}\int_{0}^{s}G(\sigma{\bf
b}_{s}-\sigma{\bf b}_{s^{\prime}})dB^{a}_{s^{\prime}}.
\end{array}\end{equation} We expect that the effect of the random metric will be seen
by an observation of the  mean distance $\vert {\bf X}-{\bf
Y}\vert^{2k} $ (as the ${\bf X}$ coordinates  are coupled to the
metric). We are interested in calculation of the average

\begin{equation}
\int  d{\bf y}d{\bf Y}\Big<P_{t}({\bf x},{\bf X};{\bf y},{\bf
Y})\vert {\bf X}-{\bf Y}\vert^{2k}\Big>
\end{equation}
 over the metric.

We use the representation ($k$ is a natural number)
\begin{equation}
\vert {\bf X}-{\bf Y}\vert^{2k}\exp(i{\bf P}({\bf X}-{\bf
Y}))=(-1)^{k}\Big(\frac{\partial^{2}}{\partial P^{j}\partial
P^{j}}\Big)^{k}\exp(i{\bf P}({\bf X}-{\bf Y}))
\end{equation}
in order to insert $ \vert {\bf X}-{\bf Y}\vert^{2k}$ in eq.(16).
Then, integrating by parts in eq.(16) (with $\psi$ depending only
on ${\bf Y}$) we obtain the result
\begin{equation}\begin{array}{l} \int d{\bf y}d{\bf Y}\Big<K^{R}_{t}({\bf x},{\bf X};{\bf
    y},{\bf Y})\Big>\vert {\bf X}-{\bf Y}\vert^{2k}\psi({\bf Y})=
    \int d{\bf
    P}\exp(-i{\bf
    P}{\bf X}) \tilde{\psi}({\bf P})(-1)^{k}\cr\Big(\frac{\partial^{2}}{\partial P^{j}\partial
P^{j}}\Big)^{k} \mathbb{E}\Big[\exp\Big(-i\sigma {\bf P}^{a}{\bf
B}_{t}^{a}-\frac{\sigma^{2}}{2} {\bf
    P}^{2}L_{t}\Big)\Big],
\end{array}\end{equation}

As $L_{t}$ is not positive definite the integral (20)
 may not exist if $\tilde{\psi}({\bf P})$ does not decay fast ( faster
 than $\exp(-\alpha {\bf P}^{2})$). We are interested in
 the limit $\psi\rightarrow 1$ (with a properly chosen topology in the space of functions $\psi$)
 \begin{equation}\begin{array}{l}\int d{\bf y}d{\bf Y}\Big<K^{R}_{t}({\bf x},{\bf X};{\bf
    y},{\bf Y})\Big>\vert {\bf X}-{\bf Y}\vert^{2k}\equiv
    \lim_{\psi\rightarrow 1} \int d{\bf y}d{\bf Y}\Big<K^{R}_{t}({\bf x},{\bf X};{\bf
    y},{\bf Y})\Big>\vert {\bf X}-{\bf Y}\vert^{2k}\psi({\bf Y})\cr
    =\int d{\bf
    P}\delta({\bf P})
    \mathbb{E}\Big[(-1)^{k}\Big(\frac{\partial^{2}}{\partial P^{j}\partial
P^{j}}\Big)^{k}\exp\Big(-i\sigma {\bf P}^{a}{\bf
B}_{t}^{a}-\frac{\sigma^{2}}{2} {\bf
    P}^{2}L_{t}\Big)\Big].
\end{array}\end{equation}For  the limit $\tilde{\psi}({\bf P})\rightarrow\delta({\bf P})$
in eq.(21) we may apply  the sequence (with $\delta\rightarrow 0$)

$\tilde{\psi}_{\delta}=\Big(\int d{\bf
P}\exp(-\frac{1}{\delta}\vert{\bf
P}\vert^{4})\Big)^{-1}\exp(-\frac{1}{\delta}\vert{\bf
P}\vert^{4})$.

Let us note the identity for ${\bf B}_{s}$ and ${\bf b}_{s}$ (in
the sense that both sides have the same probability law)
\begin{equation}
{\bf B}_{s}=\sqrt{t}{\bf B}_{\frac{s}{t}}.
\end{equation}
 We assume a scale invariant correlation function for the metric
\begin{equation}
G({\bf x})=\vert {\bf x}\vert^{-2\gamma}.
\end{equation}
As discussed at eq.(8) such a scale invariant  correlation
function appears in quantum gravity . Using the scale invariant
$G$ we obtain an exact dependence of $L_{t}$ on $t$

\begin{equation}\begin{array}{l}
L_{t}=2t^{1-\gamma}\kappa\int_{0}^{1}dB_{s}^{a}\int_{0}^{s}G(\sigma{\bf
b}_{s}-\sigma{\bf b}_{s^{\prime}})dB^{a}_{s^{\prime}}\equiv
t^{1-\gamma}L_{1}.
\end{array}\end{equation}We are interested only in the
short time behaviour of the process ${\bf X}_{t}$. For such
estimates it would be sufficient to assume that the behaviour (23)
holds true at small distances in order to show that the formula
(24) holds true for a small time with a negligible remainder. For
the operator $\tilde{A}$ in the kernel of eq.(16) ${\bf
x}_{t}({\bf x})$ would be a non-Gaussian stochastic process with
the drift $\nabla_{{\bf x}}\ln \det g({\bf x}_{s})$ discussed at
the beginning of this section (eq.(1)). Then, in eq.(17) we would
have $\sigma{\bf x}_{s}-\sigma{\bf x}_{s^{\prime}}$ instead of
$\sigma{\bf b}_{s}-\sigma{\bf b}_{s^{\prime}}$. In order to obtain
the estimate (24) with a small remainder ( required for our final
result) we would need to show that the process ${\bf x}_{s}({\bf
x})-{\bf x}$ scales as the Brownian motion (22) for a small time.
$L_{t}$ in eq.(24) plays the role of a new time. In
\cite{gikhman}(chapter 3, sec.5) the diffusion matrix is applied
in general for a random time change in order to replace $e({\bf
x}_{t})d{\bf B}$ by $d{\bf B}(\tau)$, where $\tau$ is a random
time.  A random time change is discussed in
\cite{metzler3}\cite{klafter} in order to generate processes with
anomalous diffusion.

 By means of the
normal ordering (15) we have removed from eq.(9) the infinite term
${\bf P}^{2}G(0)t$ which would describe the standard  behaviour of
the Brownian motion. We obtain $L_{t}$ as the new time variable (a
random time change). According to eq.(24) this new time variable
behaves as $t^{1-\gamma}$. For a small time this random time is
dominating the one resulting from ${\bf B}_{t}^{2}\simeq t$. This
is an intuitive explanation of the behaviour of the fourth moment
of the process (4) which  leads to $t^{2(1-\gamma)}$ replacing
$t^{2}$ of ${\bf B}_{t}^{4}$.

 For the second moment we still have (as follows from eqs.(20) and
(22)) that
\begin{equation}\begin{array}{l}
\int d{\bf y}d{\bf Y} \Big<K^{R}_{t}({\bf x},{\bf X};{\bf y},{\bf
Y})\vert {\bf X}-{\bf Y}\vert^{2}\Big> =\sigma^{2}td,
\end{array}\end{equation}
because $\mathbb{E}[{\bf B}_{t}^{2}]=td$ plays the role of time as
a consequence of the vanishing of the expectation value of the
random time\begin{equation}\mathbb{E} [L_{1}]=0.
\end{equation}

 However, for $k=2$  the random time $ t^{1-\gamma}L_{1} $ will be dominating
 for a small $t$. We have

\begin{equation}\begin{array}{l} \int d{\bf y}d{\bf Y}\Big<{\cal K}^{R}_{t}({\bf x},{\bf X};{\bf
    y},{\bf Y})\Big>\vert {\bf Y}-{\bf X}\vert^{4}
    \cr=\int d{\bf P}\delta({\bf P})\Big(\frac{\partial^{2}}{\partial P^{j}\partial
P^{j}}\Big)^{2}\mathbb{E}\Big[\exp\Big(-i{\bf P}\sigma {\bf
B}_{t}-\frac{\sigma^{2}}{2} {\bf
    P}^{2}L_{t}\Big)\Big].
    \end{array}\end{equation}
According to eq.(24)  to estimate the rhs of eq.(27) it is
sufficient to estimate $\mathbb{E}[{\bf B}_{1}^{2}L_{1}]$ and
$\mathbb{E}[L_{1}^{2}]$. The first expectation value is bounded by
the second one and by $\mathbb{E}[{\bf B}_{1}^{4}]$ on the basis
of the Schwarz inequality. Using $\mathbb{E}[(\int
fdB_{s})^{2}]=\mathbb{E}[\int f^{2}ds]$) we obtain

\begin{equation}\begin{array}{l}
\mathbb{E}[L_{1}^{2}]=4\sigma^{-4\gamma}\kappa^{2}\mathbb{E}\Big[\int_{0}^{1}ds\Big(\int_{0}^{s}
dB_{s^{\prime}} \vert {\bf b}_{s}-{\bf
b}_{s^{\prime}}\vert^{-2\gamma}\Big)^{2}\Big]
\cr=4\sigma^{-4\gamma}\kappa^{2}\int_{0}^{1}ds\int_{0}^{s}ds^{\prime}
\int d{\bf x}(2\pi)^{
    -\frac{n}{2}}\exp\Big(-\frac{{\bf x}^{2}}{2(s-s^{\prime})}\Big)
    \vert {\bf x}\vert^{-4\gamma}(s-s^{\prime})^{-\frac{n}{2}}.
\end{array}\end{equation}
We have
\begin{equation}
\int d{\bf x}(2\pi)^{
    -\frac{n}{2}}\exp\Big(-\frac{{\bf x}^{2}}{2\vert s-s^{\prime}\vert}\Big)
    \vert {\bf x}\vert^{-4\gamma}=C\vert s-s^{\prime}\vert^{-2\gamma+\frac{n}{2}}
\end{equation}
with a certain numerical constant $C$ (for convergence of eq.(29)
we need  $4\gamma<n$). Inserting (29) in eq.(28) we can see that
the integral (28) is convergent if $2\gamma<1 $ ( so $n\geq 2$).
Hence, for $2\gamma<1$ we have

\begin{equation}\begin{array}{l} \int d{\bf y}d{\bf Y}\Big<{\cal K}^{R}_{t}({\bf x},{\bf X};{\bf
    y},{\bf Y})\Big>\vert {\bf Y}-{\bf X}\vert^{4}
    \cr
    =C_{1}\sigma^{4}t^{2}+C_{2}\kappa\sigma^{4-2\gamma}t^{2-\gamma}+C_{3}\sigma^{4-4\gamma}\kappa^{2}t^{2(1-\gamma)},
    \end{array}\end{equation}
where $t^{2}$ comes from $\mathbb{E}[{\bf B}_{t}^{4}]$,
$t^{2-\gamma}$
    from $\mathbb{E}[{\bf B}_{t}^{2}L_{t}]$ and $t^{2(1-\gamma)}$ from $\mathbb{E}[{L_{t}^{2}}]$.
$C_{j}$ are numerical constants
    independent of $\sigma$ and $\kappa$.
We can see that for a small time the last term in eq.(30) (which
depends on the metric covariance)  is dominating.

We could estimate the $2k$-th power of the distance using the
estimate \cite{gikhman}
\begin{displaymath}
\mathbb{E}[(\int_{0}^{t} fdB_{s})^{2k}]\leq
c_{k}\mathbb{E}[\int_{0}^{t} f^{2k}ds]
\end{displaymath}with certain constants $c_{k}$.
It follows that (18) is finite if $k\gamma<1$.

\section{Discussion}
We have shown that the mean fourth power of the distance achieved
by a diffusing particle in a random singular metric behaves as
$t^{2(1-\gamma)}$ for a small time (depending on the random metric
singularity). Then, the index $1-\gamma$ in the heat kernel
appears in the kernel of ${\cal
A}^{-1}=\int_{0}^{\infty}dt\exp(t{\cal A})$ in $n+d$ dimensions.
The behaviour of this kernel at short distances can be related to
the reduced dimensionality $d(1-\gamma)$ of $\mathbb{R}^{d}$
\cite{jurk}\cite{carlip}\cite{habaplb}. Concerning the diffusion
on irregular structures \cite{dunne}\cite{proc} the result shows
that a singular random diffusivity may have a similar effect as a
fractal nature of the medium in which the diffusion takes place.
In the literature it is the anomalous behaviour of diffusion at
large time which is of the most interest
\cite{new}\cite{metzler3}. It can happen in models with a random
drift \cite{bouchaud}\cite{komor}. The short time behaviour is
more stable. It can appear in random walk models with a modified
random waiting time \cite{klafter} and in scale invariant models
of a diffusion on fractals. It seems that in general models with a
random diffusion generator the anomalous diffusion at small time
is possible  if the diffusion matrix is a singular random field.

 We think that the result could be
generalized to an arbitrary Riemannian manifold (in particular for
the generator $\tilde{{\cal A}}$, eq.(1)). We were concerned with
a small time $t$. In such a case only local coordinate
neighborhood is relevant in an estimate of the heat kernel.
Hopefully, the argument could be extended to an arbitrary
Riemannian metric. In principle, the mean $2k$-th power of the
distance could be measured in experiments as a confirmation of the
randomness of the metric.

Concerning  the spectral dimension of
refs.\cite{jurk}\cite{hor}\cite{dunne}\cite{carlip} we could
formally integrate in eqs.(7)-(9)  over momenta with the result
\begin{equation}\begin{array}{l} \Big<K^{R}_{t}({\bf x},{\bf X};{\bf
x},{\bf X})\Big>=(2\pi\sigma^{2})^{-\frac{n+d}{2}}
t^{-\frac{n}{2}}t^{-\frac{d}{2}(1-\gamma)}\mathbb{E}\Big[(L_{1})^{-\frac{d}{2}}\Big]
\end{array}\end{equation}
The spectral  dimension $\nu$ of the diffusion kernel defined  in
\cite{jurk}\cite{hor} when applied to eq.(31) gives
\begin{equation}\begin{array}{l}
\nu=-2\frac{d}{d\tau}\ln\Big<K^{R}_{t}({\bf x},{\bf X};{\bf
x},{\bf X})\Big>=n+d(1-\gamma)
\end{array}\end{equation} where $\tau=\ln t$.
The final result (32) does not depend on $L_{1}$. However, the
formula (31) for the diagonal of the kernel (16) makes sense only
if the last term in eq.(31) (the expectation value) is finite what
does not seem to be true because $L_{1}$ takes arbitrarily small
values. We may expect that some numerical schemes with a proper
regularization (discretization) can give a finite result for the
spectral dimension (32). However, it is unlikely that the diagonal
of the averaged heat kernel is finite with a singular random
metric.

We would encounter a difficulty if we wished to define the quantum
field theory propagator by an integration over the (proper) time
in eq.(16). For the definition of the propagator we could apply
the evolution operator $\exp(it:{\cal A}:)$. Then, we would  have
no problem with the non-positivity of ($-:{\cal A}:$). We could
also use the Gaussian metric $g^{\mu\nu}$ without any positivity
property. Instead of the rigorous Euclidean functional integral we
would have to apply the Feynman integral. The scaling properties
applied in the derivation of the modified short distance behaviour
(like in eqs.(27)-(28)) apply to the Feynman integral as well. The
difficulty  consists in establishing  the argument with the
Feynman integral as a rigorous proof.

{\bf Acknowledgement}: Co-financed from the research and
researchcommercialization fund of the University of Wroclaw

{\bf Data availability statement}: data are available on request
from the author Zbigniew Haba at zbigniew.haba@uwr.edu.pl


\begin{thebibliography}{99}
\bibitem{new}V. Sposini, A.V.Chechkin, F.Seno, G. Pagnini

 and R. Metzler,

  Random diffusivity from stochastic equations:comparison of two models
 for Brownian yet non-Gaussian diffusion,


 New Journ.Phys.{\bf 20},043044(2018)
\bibitem{metzler3} A.G. Cherstvy and R. Metzler,
Anomalous diffusion in time-fluctuating non-stationary diffusivity landscapes,

Phys.Chem.Chem.Phys.{\bf 18},23840(2016)
\bibitem{metzler4} A.V. Chechkin, F. Seno, R. Metzler and M.
Sokolov,

Brownian yet non-Gaussian diffusion:from superstatistics to subordination of diffusing diffusitivities,

Phys.Rev.{\bf X7}, 021002(2017)
\bibitem{klafter} R. Metzler and J. Klafter,

The random walk's guide to anomalous diffusion:a fractional dynamics approach,


Phys.Rep.{\bf
339},1(2000)

\bibitem{wilczek}
S. P. Robinson and F. Wilczek,

Gravitational correction to running of gauge couplings


Phys.Rev.Lett.{\bf 96},231601(2006)
\bibitem{petrykowski}A. R. Pietrykowski,

Gauge dependence of gravitational correction to running of gauge couplings,

 Phys.Rev.Lett.{\bf
98},061801(2007)
\bibitem{jurk}J. Ambjorn, J. Jurkiewicz and R.Loll,

The spectral dimension of the universe is scale dependent,


Phys.Rev.Lett.{\bf 95},173001(2005)
\bibitem{jurk2}J. Ambjorn, A. G\"orlich. J. Jurkiewicz
 and
 H.Zhang,

 The spectral dimension in 2D CDT gravity coupled to scalar fields,

 Mod.Phys.Lett. {\bf A30},1550077(2015)
\bibitem{jurk3} J. Ambjorn, K.N. Anagnostopoulos, L. Jensen,
T.Ichihara and Y. Watabiki,

Quantum geometry and diffusion,

 JHEP{\bf 11}(1998)022
\bibitem{hor}P. Horava,

Spectral dimension of the universe in quantum gravity at a Lifshitz point,

 Phys.Rev.Lett.{\bf 102},161301(2009)

\bibitem{hor2}P. Horava, Quantum gravity at a Lifshitz point,

 Phys.Rev.{\bf D79},084008(2008)
\bibitem{liou}S. Andres and N.Kajino,

Continuity of the heat kernel and spectral dimension of Liouville Brownian motion,

Prob.Theory Relat.Fields,
{\bf 166},713(2016)
\bibitem{carlip}S. Carlip,


Dimension and dimensional reduction in quantum gravity,



Class.Quant.Grav.{\bf 34},193001(2017)
\bibitem{carlip2} S. Carlip,

Spacetime foam: a review,

 Rep.Prog.Phys.2023,May 5;86(6)
\bibitem{dunne}G.V. Dunne,

Heat kernels and zeta functions on fractals,

J.Phys.{\bf A45},374016(2012)

\bibitem{proc} B. O'Shaughnessy and I. Procaccia,

Analytical solutions for diffusion on fractal objects,

   Phys.Rev.Lett.{\bf 54},455(1985)
   \bibitem{bar}M.T. Barlow, Diffusion on fractals, Lecture
   Notes in Math.{\bf 1690}(1998)

\bibitem{smolin} L. Crane and L. Smolin,

Renormalization of general relativity on a background of spacetime foam,

 Nucl.Phys.{\bf
B267},714(1986)

\bibitem{habaplb}Z.Haba,

Universal regular short distance behavior from an  interaction with a scale invariant gravity,

 Phys.Lett.{\bf B528},129(2002)
\bibitem{habajpa}Z. Haba,

The $\Phi$4 quantum  field in a scale invariant random metric,

Journ.Phys.{\bf A35},7425(2002)
\bibitem{gikhman}I.I. Gikhman and A.V. Skorohod, Stochastic
differential equations,

Springer,New York,1972

\bibitem{bouchaud}J.P. Bouchaud and A. Georges,

Anomalous diffusion in disordered media:statistical mechanisms, models and physical applications,

Phys.Rep. {\bf
195},127(1990)
\bibitem{komor}T. Komorowski and S. Olla,

On the superdiffusive behavior of passive tracer with a Gaussian drift,

 Journ.Stat.Phys.
{\bf 108},647(2002)



\bibitem{simon}B. Simon, Functional integration and quantum
physics,Academic, New York,1979
  \bibitem{berger}M.A. Berger and V.J. Mizel,

  Theorems of Fubini type for iterated stochastic integrals,

 Trans.Amer.Math.Soc.{\bf
    252},249(1979)

    \bibitem{albeverio}S. Albeverio and S. Kusuoka,

    A basic estimate for two-dimensional stochastic holonomy along Brownian bridges,


    J.Funct.Anal.{\bf 127},132(1994)



 \bibitem{betz}V. Betz and F.
    Hiroshima,


    Gibbs measures with double stochastic integrals on a path space,

    Inf.Dim.Anal.Quant.Prob. Rel.Topics,{\bf
    12},135(2009)



    \end{thebibliography}
\end{document}